\documentclass[a4paper]{article}

\usepackage[utf8]{inputenc}
\usepackage{times}
\usepackage{authblk}
\usepackage[]{algorithm, algorithmic}

\usepackage{natbib}

\usepackage{graphicx}

\usepackage[labelfont=bf, labelsep=quad]{caption}
\usepackage{subcaption}

\usepackage{amsmath}
\usepackage{amsfonts}
\usepackage{amssymb}

\DeclareMathOperator*{\argmin}{arg\,min}

\usepackage[top=1.25in, bottom=1.25in, left=1.25in, right=1.25in]{geometry}

\usepackage{multirow}

\makeatletter
\newenvironment{breakablealgorithm}
  {
   \begin{center}
     \refstepcounter{algorithm}
     \hrule height.8pt depth0pt \kern2pt
     \renewcommand{\caption}[2][\relax]{
       {\raggedright\textbf{\ALG@name~\thealgorithm} ##2\par}
       \ifx\relax##1\relax 
         \addcontentsline{loa}{algorithm}{\protect\numberline{\thealgorithm}##2}
       \else
         \addcontentsline{loa}{algorithm}{\protect\numberline{\thealgorithm}##1}
       \fi
       \kern2pt\hrule\kern2pt
     }
  }{
     \kern2pt\hrule\relax
   \end{center}
  }
\makeatother

\providecommand{\keywords}[1]
{
  \small	
  \textbf{\textit{Keywords---}} #1
  \normalsize
}

\usepackage{appendix}

\begin{document}
\title{Bayesian Boosting for Linear Mixed Models}
\author[$1$,*]{Boyao Zhang}
\author[$1$]{Colin Griesbach}
\author[$2$]{Cora Kim}
\author[$2$]{Nadia Müller-Voggel}
\author[$1$]{Elisabeth Bergherr}

\affil[$1$]{Department of Medical Informatics, Biometry and Epidemiology, University of Erlangen-Nuremberg, Waldstrasse 6, 91054 Erlangen, Germany}
\affil[$2$]{Department of Neurosurgery, University Hospital, Erlangen, Germany}

\affil[*]{Corresponding author {\sf{e-mail: boyao.zhang@fau.de}}, Phone: +49-(0)9131-85-22729, FAX: +49-(0)9131-85-25740}
\maketitle  

\begin{abstract}
Boosting methods are widely used in statistical learning to deal with high-dimensional data due to their variable selection feature.
However, those methods lack straightforward ways to construct estimators for the precision of the parameters such as variance or confidence interval, which can be achieved by conventional statistical methods like Bayesian inference.
In this paper, we propose a new inference method ``BayesBoost" that combines boosting and Bayesian for linear mixed models to make the uncertainty estimation for the random effects possible on the one hand.
On the other hand, the new method overcomes the shortcomings of Bayesian inference in giving precise and unambiguous guidelines for the selection of covariates by benefiting from boosting techniques.
The implementation of Bayesian inference leads to the randomness of model selection criteria like the conditional AIC (cAIC), so we also propose a cAIC-based model selection criteria that focus on the stabilized regions instead of the global minimum.
The effectiveness of the new approach can be observed via simulation and in a data example from the field of neurophysiology focussing on the mechanisms in the brain while listening to unpleasant sounds.
\end{abstract}

\keywords{Bayesian inference, Boosting, Linear mixed models, conditional AIC, Variable selection, Neurophysiology}

\section{Introduction}
\label{sec: intro}

Linear mixed models (LMM) \citep{laird1982} are widely used in longitudinal data analysis as they incorporate random effects to deal with group-specific heterogeneity. Data involving repeated observations of the same variables are common in epidemiology, medical statistics and many other fields.

Likelihood-based methods are often used to make inference for (generalized) linear mixed models \citep{bates2000mixed, gumedze2011}.
\citet{schelldorfer2011} and \citet{groll2014variable} introduced separately the $L1$-penalized estimation for high-dimensional linear mixed models.
\citet{fong2010} argued that for small sample sizes likelihood-based inference can be unreliable with variance components being difficult to estimate and suggested to use the Bayesian method.
When the random effects distribution is misspecified, the resulting maximum likelihood estimators are inconsistent and biased \citep{neuhaus1992, heagerty2001, litiere2008}.
\citet{fahrmeir2001bayesian} presented a fully Bayesian inference via Markov Chain Monte Carlo (MCMC) simulation in generalized additive and semiparametric mixed models.
\citet{rosa2003} described a normal/independent residual distributions for robust inference and suggested also the Bayesian framework.
Bayesian inference for mixed models can be conducted with for example BayesX, a program with MCMC simulation techniques \citep{lang2000}.

Regarding variable selection, \citet{tutz2010generalized} proposed an additional regularization approach for generalized linear mixed models using likelihood-based boosting.
Especially with focus on penalized likelihood inference, likelihood-based boosting \citep{tutz2006generalized} represents an alternative to well known gradient boosting techniques \citep{buhlmann2003, hothorn2010} and is due to its componentwise maximization routine suitable for variable selection and high-dimensional data structures.
\citet{griesbach2019addressing} addressed potential bias of the likelihood-based boosting estimators occurring in the presence of cluster-constant covariates like gender or treatment group in longitudinal studies and proposed an improved algorithm where the random effects are ensured to be uncorrelated with any observed variables.
In \citet{colin2021} this bias correction was successfully adapted to gradient boosting estimating technique for linear mixed models.
However, variable selection regarding the random structure in this approach is not allowed, as the random effects have to be specified in advance.


Model selection for mixed models is usually based on the Akaike information criterion (AIC) \citep{akaike1973}.
\citet{vaida2005} argued that the AIC is not appropriate for the focus on clusters and proposed an AIC derived from the conditional model formulation with the known variance-covariance matrix of random effects, known as the conditional AIC (cAIC).
\citet{liang2008} provided a corrected cAIC that accounts for the estimation of the variance parameters.
\citet{greven2010} proved that the marginal AIC (mAIC) is an asymptotically biased estimator of the Akaike information that favors smaller models without random effects.
Thus, they suggested also the cAIC and derived an analytic representation of the corrected version of cAIC, which saves computing time \citep{sfken2018conditional}.

While boosting provides a very flexible model-based inference, there is no straightforward way to conduct standard parametric hypothesis tests. 
The biased estimates induced by shrinkage affects also other viable alternatives such as bootstrap confidence interval \citep{hepp2019, mayr2017}.
However, with Bayesian sampling, the variability of the coefficients (e.g.\ the credible interval) is already part of the estimation procedure.
So in this paper, we introduce a new inference method that combines boosting and Bayesian for linear mixed models, denoted as ``BayesBoost", which incorporates the two concepts for the first time and benefits from the shrinkage and variable selection properties of boosting and from the uncertainty estimates of MCMC simulation.

BayesBoost divides the estimation procedure into two parts, the componentwise gradient boosting estimation for the fixed effects and the merged estimation (boosting and Gibbs sampler) for the random effects.
The automatic selection of the random effects is based on their contribution to the model measured by e.g.\ the in-sample mean squared error (MSE), and it is assumed that only informative fixed effects have random effects, i.e.\ only when a covariate is selected as a fixed effect, it can be the participant for the choice of random effects.
We also provide the interface for fixed user-defined random effects structure or flexible user-defined random effect participants set to make the automatic selection without this assumption.

To prevent overfitting, we introduce a cAIC-based model selection criteria.
Due to the possible occurrence of outliers in the MCMC procedure, the cAIC for the model in some iterations can be extremely small/large.
We thus adopt the Hampel-filter \citep{pearson1999}, which is often used for the data cleaning in time series and signal processing, that replaces the outliers detected by the median absolute deviation \citep{rousseeuw1993} with the median in a moving-window.
However, this filter cannot avoid small oscillations and waves of the cAIC series caused by the MCMC samplings.
We are inspired by the early stopping meta-algorithm \citep{goodfellow2016} from deep learning due to its ``patience" parameter that controls the number of times to observe better iteration position before stopping.
So we rewrite this algorithm to make it fit for finding the local minimum from cAIC series.

The paper is structured as follows: Section \ref{sec: method} specifies the linear mixed models and introduces how to make inference with the BayesBoost algorithm. Random effects selection and model selection are also covered in this section. Section \ref{sec: sim} evaluates the performance of the BayesBoost algorithm in two simulations and highlights its new features. Section \ref{sec: app} presents the application of BayesBoost algorithm on the cognitive neuroscience experiment about the brain reflection listening to unpleasant sounds. Section \ref{sec: summary} concludes with a discussion and outlook. A detailed BayesBoost algorithm is included in the appendix.

\section{Methods}
\label{sec: method}
This section starts with the specification of linear mixed models for longitudinal and clustered data. 
Then we propose the BayesBoost algorithm for the parameter estimation.
A cAIC-based model selection criteria is also introduced in this section.

\subsection{Model specification}
For subjects $j = 1, \dots, n_i$ in clusters or individuals $i = 1, \dots, m$ with $n = \sum_{i=1}^m n_i$, consider the linear mixed model
\begin{align*}
\boldsymbol{y}_i = \boldsymbol{X}_i \boldsymbol{\beta} + \boldsymbol{Z}_i \boldsymbol{\gamma}_i + \boldsymbol{\varepsilon}_i,
\end{align*}
where $\boldsymbol{y}_i$ is the $n_i$-dimensional vector of responses for individual $i$, $\boldsymbol{X}_i$ and $\boldsymbol{Z}_i$ are $n_i \times (p+1)$- and $n_i \times (q + 1)$-dimensional design matrices constructed from known covariates, $\boldsymbol{\beta}$ is a $(p+1)$-dimensional vector of fixed effects with intercept, $\boldsymbol{\gamma}_i$ is a $(q + 1)$-dimensional vector of cluster-specific random effects with random intercept, and $\boldsymbol{\varepsilon}_i$ is a $n_i$-dimensional vector of errors.
This model can be formulated in a rather compact form as
\begin{align}
\boldsymbol{y} &= \boldsymbol{X} \boldsymbol{\beta} + \boldsymbol{Z} \boldsymbol{\gamma} + \boldsymbol{\varepsilon}
\end{align}
with 
\begin{align}
\begin{pmatrix}
\boldsymbol{\gamma} \\
\boldsymbol{\varepsilon}
\end{pmatrix} &\sim N\begin{pmatrix}
\begin{pmatrix}
\boldsymbol{0} \\
\boldsymbol{0}
\end{pmatrix},
\begin{pmatrix}
\boldsymbol{G} & \boldsymbol{0} \\
\boldsymbol{0} & \boldsymbol{R}
\end{pmatrix}
\end{pmatrix},
\end{align}
where $\boldsymbol{y} = (\boldsymbol{y}_1, \dots, \boldsymbol{y}_m)^T$, $\boldsymbol{\varepsilon} = (\boldsymbol{\varepsilon}_1, \dots, \boldsymbol{\varepsilon}_m)^T$ and $\boldsymbol{\gamma} = (\boldsymbol{\gamma}_1, \dots, \boldsymbol{\gamma}_m)^T$, as well as the design matrices $\boldsymbol{X} = (\boldsymbol{X}_1, \dots, \boldsymbol{X}_m)^T$ and $\boldsymbol{Z} = \text{blockdiag}(\boldsymbol{Z}_1, \dots, \boldsymbol{Z}_m)$.
The covariance matrices for $\boldsymbol{\gamma}$ and $\boldsymbol{\varepsilon}$ are assumed to be positive definite, and $\boldsymbol{\gamma}$ and $\boldsymbol{\varepsilon}$ are independent.
The covariance $\boldsymbol{G}$ and $\boldsymbol{R}$ are block-diagonal matrices with
\begin{align*}
\boldsymbol{R} &= \text{blockdiag}(\sigma^2 \boldsymbol{\Sigma}_{n_1}, \cdots, \sigma^2 \boldsymbol{\Sigma}_{n_i}, \cdots, \sigma^2 \boldsymbol{\Sigma}_{n_m}) \\
\boldsymbol{G} &= \text{blockdiag}(\boldsymbol{Q}, \cdots, \boldsymbol{Q}, \cdots, \boldsymbol{Q}),
\end{align*}
where $\boldsymbol{\gamma}_i \sim N(\boldsymbol{0}, \boldsymbol{Q})$ with $(1+q) \times (1+q)$-covariance matrix $\boldsymbol{Q}$. 
For i.i.d.\ errors, which is also the case in this paper, $\boldsymbol{R}$ simplifies to $\boldsymbol{R} = \sigma^2 \boldsymbol{I}$.

\subsection{BayesBoost}
From a BayesBoost perspective, only random effects can be considered as random variables, while fixed effects are assumed to be constant.
So unlike in conventional full Bayesian inference, where the fixed effects vector $\boldsymbol{\beta}$ is obtained by MCMC simulation, it is updated by a boosting step in each iteration of the BayesBoost approach. 
Samples of other random effects parameters are then drawn according to their full conditional distributions with respect to the estimated $\hat{\boldsymbol{\beta}}$ as the usual MCMC procedure.

Specifically, the predictor $\boldsymbol{\eta}$ for the response displays as 
\begin{align*}
\boldsymbol{y} = \boldsymbol{\eta} = \sum_{k=1}^p \boldsymbol{\eta}_k
\end{align*}
with
\begin{align}
\boldsymbol{\eta}_k = \boldsymbol{X}_k \boldsymbol{\beta}_k + \boldsymbol{Z}_k \boldsymbol{\gamma}_k,
\label{eq: predictor}
\end{align}
where $\boldsymbol{X}_k = (1, X_k)$ is a design matrix that accounts for the intercept and the corresponding parameter vector is $\boldsymbol{\beta}_k = (\beta_0, \beta_k)$.
Representation \eqref{eq: predictor} divides  the predictor into fixed and random parts, and the fixed part should be estimated first in a boosting iteration. The estimation of the separate parts will be explained in the following two paragraphs.

\subsubsection{Fixed effects estimation}
In componentwise gradient boosting, the negative gradient vector
\begin{align*}
\boldsymbol{u}^{[s]} = \frac{\partial \rho(\boldsymbol{y}, \hat{\boldsymbol{\eta}}^{[s-1]})}{\partial \boldsymbol{\eta}} = \boldsymbol{y} - \hat{\boldsymbol{\eta}}^{[s-1]}, \quad s = 1, \dots, m_{\text{stop}}
\end{align*}
with the L2-loss $\rho(\cdot)$, i.e.\ $\rho(a, b) = \frac{1}{2}\sum (a-b)^2$, in boosting iteration $s$ is fitted to each base-learner $h(X_k), k = 1, \dots, p$ separately.
The best-fitting base-learner is then selected based on the residual sum of squares with respect to $\boldsymbol{u}^{[s]}$
\begin{align*}
k^* = \argmin_{k \in \{1, \dots, p\}} \sum_{i = 1}^n (u_i^{[s]} - \hat{h}(X_{ik}))^2,
\end{align*}
where $X_{ik}$ denotes the $i$-th observation of $X_k$.
The updated coefficients of fixed effects with respect to the best-fitting covariate are
\begin{align}
\hat{\boldsymbol{\beta}}^{[s]} = \hat{\boldsymbol{\beta}}^{[s-1]} + \nu \hat{\boldsymbol{\beta}}_{k^*}^{[s]},
\label{eq: update_beta}
\end{align}
where $\nu$ denotes a step-length or learning rate.
Note that $\hat{\boldsymbol{\beta}}^{[s-1]}$ is a vector of length $(p+1)$ and $\hat{\boldsymbol{\beta}}_{k^*}^{[s]}$ is of length $2$,
so updates of $\hat{\boldsymbol{\beta}}^{[s]}$ in equation \eqref{eq: update_beta} happens only in the intercept term and the corresponding $k^*$-th covariable, while all the other covariates remains unchanged.

Model improvement benefiting from the fixed effect $X_{k^*}$ is measured with the MSE
\begin{align}
\text{MSE}_{k^*, \text{fixed}} = \frac{1}{n}\sum_{i=1}^n \left(u_i - \boldsymbol{X}_{ik^*} \hat{\boldsymbol{\beta}}_{k^*}^{[s]} \right)^2.
\label{eq: mse_fix}
\end{align}
This serves later for deciding whether $X_{k^*}$ should also contribute to the model as a random effect, i.e.\ the selection of random effects.

\subsubsection{Random effects estimation}
From the conventional Bayesian perspective, both fixed effects $\boldsymbol{\beta}$ and random effects $\boldsymbol{\gamma}$ are considered to be random variables.
In BayesBoost, however, fixed effects have already been estimated in the boosting step as constants and should not be treated as random variables any more.
Therefore, the full Bayesian inference for the parameters of interest is based on the posterior distribution
\begin{align}
p(\boldsymbol{\gamma}, \boldsymbol{G}, \boldsymbol{R} \vert \tilde{\boldsymbol{y}}) \propto p(\tilde{\boldsymbol{y}} \vert \boldsymbol{\gamma}, \boldsymbol{G}, \boldsymbol{R}) p(\boldsymbol{\gamma} \vert \boldsymbol{G}) p(\boldsymbol{G}) p(\boldsymbol{R}),
\label{eq: poster_dist}
\end{align}
with $\tilde{\boldsymbol{y}} = \boldsymbol{y} - \boldsymbol{X} \hat{\boldsymbol{\beta}}^{[s]}$ if we treat the fixed effects as an offset term, and $\boldsymbol{\gamma} \vert \boldsymbol{G}$, $\boldsymbol{R}$ and $\boldsymbol{G}$ are assumed to be independent.
In general, \eqref{eq: poster_dist} cannot be displayed in a closed form, such that the full Bayesian inference is usually conducted through MCMC simulation.

The random effects distribution $\boldsymbol{\gamma} \sim N(\boldsymbol{0}, \boldsymbol{G})$ can be seen as a prior for the random effects.
The full conditional distribution of $\boldsymbol{\gamma}$ is then
\begin{align}
p(\boldsymbol{\gamma} \vert \tilde{\boldsymbol{y}}, \boldsymbol{G}, \boldsymbol{R}) \propto& p(\tilde{\boldsymbol{y}} \vert \boldsymbol{\gamma}, \boldsymbol{G}, \boldsymbol{R}) p(\boldsymbol{\gamma} \vert \boldsymbol{G}) \nonumber\\
\propto& \exp \left(-\frac{1}{2} \left(\tilde{\boldsymbol{y}} - \boldsymbol{Z} \boldsymbol{\gamma}\right)^T \boldsymbol{R}^{-1} \left(\tilde{\boldsymbol{y}} - \boldsymbol{Z} \boldsymbol{\gamma} \right) \right) \exp\left(-\frac{1}{2} \boldsymbol{\gamma}^T \boldsymbol{G}^{-1} \boldsymbol{\gamma} \right).
\label{eq: p_gamma}
\end{align}
It can be shown that \eqref{eq: p_gamma} forms a Gaussian distribution with parameters
\begin{align*}
\boldsymbol{\Sigma}_{\boldsymbol{\gamma}} &= \left(\boldsymbol{Z}^T \boldsymbol{R}^{-1} \boldsymbol{Z} + \boldsymbol{G}^{-1} \right)^{-1}, \\
\boldsymbol{\mu}_{\boldsymbol{\gamma}} &= \boldsymbol{\Sigma}_{\boldsymbol{\gamma}} \left( \boldsymbol{Z} \boldsymbol{R}^{-1} \left(\boldsymbol{y} - \boldsymbol{X} \hat{\boldsymbol{\beta}}^{[s]} \right) \right).
\end{align*}

According to the model specification, the covariance matrix for i.i.d.\ errors $\boldsymbol{R} = \sigma^2 \boldsymbol{I}$ is dominated by the hyperparameter $\sigma^2$, and a weakly informative inverse gamma prior $\sigma^2 \sim IG(a, b)$ with small $a$ and $b$ is commonly proposed, because for $a = b$ and both values approaching zero, the distribution of $\log \sigma^2$ tends to be a uniform distribution.
Thus small values for $a$ and $b$ are identified with a weakly informative or noninformative prior.
The fully conditional density of $\sigma^2$ turns out to be
\begin{align*}
p(\sigma^2 \vert \tilde{\boldsymbol{y}}, \boldsymbol{\gamma}) &\propto p(\tilde{\boldsymbol{y}} \vert \boldsymbol{\gamma}, \sigma^2) p(\sigma^2) \\
&\propto (\sigma^2)^{-\frac{n}{2}} \exp \left(-\frac{1}{2 \sigma^2}(\tilde{\boldsymbol{y}} - \boldsymbol{Z} \boldsymbol{\gamma})^T (\tilde{\boldsymbol{y}} - \boldsymbol{Z} \boldsymbol{\gamma}) \right) \cdot (\sigma^2)^{-a-1} \exp \left(-\frac{1}{\sigma^2} b \right).
\end{align*}
It is again an inverse gamma distribution $IG(\tilde{a}, \tilde{b})$ with
\begin{align*}
\tilde{a} &= a + \frac{n}{2}, \\
\tilde{b} &= b + \frac{1}{2} \left(\boldsymbol{y} - \boldsymbol{X} \hat{\boldsymbol{\beta}}^{[s]} - \boldsymbol{Z} \boldsymbol{\gamma} \right)^T \left(\boldsymbol{y} - \boldsymbol{X} \hat{\boldsymbol{\beta}}^{[s]} - \boldsymbol{Z} \boldsymbol{\gamma} \right).
\end{align*}

The last parameter whose prior needs to be specified is the covariance matrix $\boldsymbol{G}$ of the random effects.
Analogously, the block-diagonal matrix $\boldsymbol{G} = \text{blockdiag}(\boldsymbol{Q}, \dots, \boldsymbol{Q})$ is dominated by the covariance matrix of single cluster or individual $\boldsymbol{Q}$.
Usually, we assume an inverse Wishart prior for the covariance matrix, i.e.\ $\boldsymbol{Q} \sim IW(v_0, \boldsymbol{\Lambda}_0)$, which can be understood as the multivariate case of an inverse gamma distribution.
Recalling that $\boldsymbol{Q}$ is a $(1+q) \times (1+q)$-dimensional matrix and there exists totally $m$ clusters, we have
\begin{align*}
p(\boldsymbol{Q} \vert \tilde{\boldsymbol{y}}, \boldsymbol{\gamma}, \boldsymbol{R}) &\propto p(\tilde{\boldsymbol{y}} \vert \boldsymbol{\gamma}, \boldsymbol{Q}, \boldsymbol{R}) p(\boldsymbol{\gamma} \vert \boldsymbol{Q}) p(\boldsymbol{Q}) \\
&\propto p(\boldsymbol{\gamma} \vert \boldsymbol{Q}) p(\boldsymbol{Q}) \\
&\propto |\boldsymbol{Q}|^{-\frac{m}{2}} \exp\left(-\frac{1}{2} \sum_{i=1}^m \boldsymbol{\gamma}_i^T \boldsymbol{Q}^{-1} \boldsymbol{\gamma}_i \right) \cdot |\boldsymbol{Q}|^{-\frac{v_0 + (1+q) + 1}{2}} \exp\left(-\frac{1}{2} \text{tr}\left(\boldsymbol{Q}^{-1} \boldsymbol{\Lambda}_0 \right) \right),
\end{align*}
where $|\cdot|$ denotes the determinant of a matrix. 
Therefore, the full conditionals for $\boldsymbol{Q}$ is an inverse Wishart distribution with
\begin{align*}
v &= v_0 + m, \\
\boldsymbol{\Lambda} &= \boldsymbol{\Lambda}_0 + \boldsymbol{\gamma}^T \boldsymbol{\gamma}.
\end{align*}

The corresponding Gibbs samples can then be drawn from the full conditional distributions for unknown parameters.
In conventional Bayesian inference for linear mixed models, the MCMC simulation procedure is performed only once.
However, the combination of Bayesian inference and boosting makes it necessary that the procedure is performed the same number of times as the total of boosting iterations since the estimated values for the parameters are required at the end of every iteration.
To minimize the invalid samples (burn-in) to the most extent, we can use the posterior modes from the last iteration instead of repeatedly using the initialized values as the starting values of the MCMC process for the current iteration.
The preference to the mode instead of mean is due to the asymmetric prior distribution of $\sigma^2$ and $\boldsymbol{Q}$, and the sample outliers affect the mean but have little effect on the mode.
We omit burn-in of the individual MCMC chains for two reasons: the first one is that, due to the gradual convergence to the correct values the starting values in the later boosting iterations are already in the correct area, since they are based on the results of the previous boosting iteration. 
Hence, we could consider the whole chain generated by all boosting iterations except the last as pseudo burn-in. 
The second reason is the stability we get from the parameters generated solely by the boosting mechanism: the full conditional distributions are derived on constant fixed effects $\hat{\boldsymbol{\beta}}^{[s]}$, which also leads to less instability than seen in full MCMC approaches. 
There will not exist large changes in the burn-in period between the samples as seen in conventional Bayesian inference, where $\boldsymbol{\beta}$ is regarded a random variable and must be drawn as well because the process of approaching the stationary region is omitted.

Another point deserving to be mentioned here is the nearest positive definite matrix.
Model selection discussed in this paper is based on the conditional AIC, which requires a Cholesky decomposition of the covariance matrix $\boldsymbol{G}$ or just $\boldsymbol{Q}$ to avoid calculating the high-dimensional inverse matrix \citep{sfken2018conditional}.
The covariance matrix $\boldsymbol{Q}$ constructed from the elementwise posterior mode of Gibbs samples, however, does not guarantee to be a positive definite matrix, i.e.\ the condition of Cholesky decomposition is not fulfilled.
In practice, for the case of non-positive definite $\boldsymbol{Q}$, we suggest to transform it to its nearest positive definite matrix.
The transforming algorithm is beyond the scope of this paper, for more details please refer to \citet{higham2002computing}.

\subsubsection{Bias correction}
A common problem of the likelihood-based boosting approach is the correlation between random intercepts and cluster-constant covariates (e.g.\ gender or age-group), and \citet{griesbach2019addressing} addressed the problem by introducing an additional correction step for the random effects within a usual boosting framework.
But unlike the ``piecewise" updates of random effects in the usual boosting algorithms, BayesBoost extracts the whole information of random effects from the residuals (i.e.\ $\boldsymbol{y} - \boldsymbol{X}\hat{\boldsymbol{\beta}}^{[s]}$) at once in every boosting iteration through MCMC simulation.
This mechanism often induces ineffective updates of fixed effects.
Due to stepwise build up of the boosting approach the fixed effects explain only a little variance of the response in the beginning iterations, such that the residuals contain lots of information, which ought to belong to the fixed effects part, but are explained by the random effects altogether.
Consequently, the information extracted by the fixed effects in the next boosting iterations from the remaining residuals $\boldsymbol{y} - \boldsymbol{X} \hat{\boldsymbol{\beta}}^{[s-1]} - \boldsymbol{Z} \hat{\boldsymbol{\gamma}}^{[s-1]}$ becomes ineffective since lots of information have already been accounted by the random part.

Therefore, weakened and disentangled updates are meaningful not only for the random intercepts but also for the random slopes in the BayesBoost framework.
To prevent such correlation between fixed and random effects, we replace the original design-matrix of the random effects $\tilde{\boldsymbol{Z}}$ with
\begin{align}
\boldsymbol{Z} = (\boldsymbol{I} - \boldsymbol{X} (\boldsymbol{X}^T \boldsymbol{X})^{-1} \boldsymbol{X}^T) \tilde{\boldsymbol{Z}},
\label{eq: Z}
\end{align}
where $\boldsymbol{I} - \boldsymbol{X} (\boldsymbol{X}^T \boldsymbol{X})^{-1} \boldsymbol{X}^T$ is known as the residual maker matrix, which is interpreted as a matrix that produces the least squares residuals in the regression of $\tilde{\boldsymbol{Z}}$ on $\boldsymbol{X}$ when it multiplies any $\tilde{\boldsymbol{Z}}$.
Note that the design matrix $\boldsymbol{Z}$ used in the entire algorithm including what we have discussed above is actually the corrected version \eqref{eq: Z}.
Based on the linear regression theory it can be easily proved that $\boldsymbol{X}^T\boldsymbol{Z} = 0$, i.e.\ $\boldsymbol{Z}$ is uncorrelated with $\boldsymbol{X}$.
From this perspective, the corrected design matrix $\boldsymbol{Z}$ and $\boldsymbol{X}$ creates two-dimensional orthogonal subspaces.
Variations or updates of random effects on the $\boldsymbol{Z}$-subspace will not influence the fixed effects on the $\boldsymbol{X}$-subspace due to this orthogonal projection.
This transformation helps limiting the explanation scope of random effects, i.e.\ the random effects can only explain the variation of response that cannot be explained by the fixed effects.

\subsubsection{Random effects selection}
The last step of the algorithm is the selection of random effects.
To make it possible, the covariance structure of random effects needs to be reconstructed at the sampling step.
Based on the assumption that only fixed effects can have random effects, as long as the best-fitting fixed effect $X_{k^*}$ in iteration $s$ has not served as a random effect, a temporary or potential covariance matrix $\boldsymbol{Q}_{\text{pot}}^{[s]}$ shall be constructed
\begin{align*}
\boldsymbol{Q}_{\text{pot}}^{[s]} = \text{diag}(\boldsymbol{Q}^{[s-1]}, 1),
\end{align*}
with $1$ as the initialized starting value.
The change of the covariance structure affects also other relevant elements for example the hyperparameter $\boldsymbol{\Lambda}_{0, \text{pot}}^{[s]}$, and the design matrix $\boldsymbol{Z}_{\text{pot}}^{[s]}$ should also coincide with the new structure.
Estimates, especially $\boldsymbol{\gamma}_{\text{mode, pot}}^{[s]}$ (posterior mode), based on the new structure enables us to have access to the model improvement benefiting from $X_{k^*}$ as both a fixed and a random effect, i.e.\
\begin{align}
\text{MSE}_{k^*} = \frac{1}{n}\sum_{i=1}^n \left(u_i - \boldsymbol{X}_{ik^*} \hat{\boldsymbol{\beta}}_{k^*}^{[s]} - \boldsymbol{Z}_{i k^*, \text{pot}}^{[s]} \hat{\boldsymbol{\gamma}}_{k^*, \text{mode, pot}}^{[s]} \right)^2,
\label{eq: mse_random}
\end{align}
where $\boldsymbol{Z}_{k^*, \text{pot}}^{[s]}$ denotes the submatrix of $\boldsymbol{Z}_{\text{pot}}^{[s]}$ accounting for the $k^*$-th covariate.

Decisions on the selection of random effects can thus be made by comparing the mean squared error in terms of the $X_{k^*}$ as a fixed effect in \eqref{eq: mse_fix} and that as both fixed and random in \eqref{eq: mse_random}.
If the model improvement of $X_{k^*}$ as both fixed and random effect is greater than as only fixed, the structures of potential covariance as well as other elements shall be held, otherwise they ought to be reset to the previous status.

Theoretically, if $X_{k^*}$ is not additionally recognized as a random effect, the MCMC simulation process should be conducted again with the old covariance structure to get more precise estimations.
But we can still keep the estimates excluding only $X_{k^*}$ relevant values practically (e.g.\ keep the already selected random effects in the previous iterations $\hat{\boldsymbol{\gamma}}_{0, \text{mode}}^{[s]}$ and $\hat{\boldsymbol{\gamma}}_{1, \text{mode}}^{[s]}$ from $\hat{\boldsymbol{\gamma}}_{\text{mode, pot}}^{[s]} = \left(\hat{\boldsymbol{\gamma}}_{0, \text{mode}}^{[s]}, \hat{\boldsymbol{\gamma}}_{1, \text{mode}}^{[s]}, \hat{\boldsymbol{\gamma}}_{k^*, \text{mode}}^{[s]}\right)^T$ as the estimate for $\hat{\boldsymbol{\gamma}}^{[s]}$), because a larger $\text{MSE}_{k^*}$ implies also small influence of $X_{k^*}$ on the model.
In other words, the estimates within the old covariance structure have already explained the model well, so ignoring the $X_{k^*}$ relevant estimates will have little effect.
And for the efficiency concerns, a second MCMC procedure is a huge computing burden.
Therefore, in our proposed algorithm, the estimates sampled from the potential structures are still updated even if $X_{k^*}$ is not selected as a random effect.

Finally, we present a sketch of BayesBoost in Algorithm \ref{alg: sketch} that summarizing its basic mechanism and ideas. 
A detailed description of BayesBoost is given at Appendix \ref{apx: bayesboost}.

\begin{algorithm}[ht]
\caption{BayesBoost for linear mixed models}
\label{alg: sketch}
\begin{algorithmic}[1]
\STATE Initialization.
\STATE Calculate the uncorrelated design-matrix $\boldsymbol{Z}$.
\FOR{Boosting iteration $s = 1$ to $m_{\text{stop}}$}
	\STATE Find the best-fitting variable $X_{k^*}, k^* \in \{1, \dots, p\}$ and its coefficient $\hat{\boldsymbol{\beta}}_{k^*}$ (with intercept) by fitting componentwise regression models to the pseudo-residuals.
	\IF{$X_{k^*}$ serves not as a random effect in iteration $s-1$} \label{alg: sketch5}
		\STATE Construct potential covariance matrix $\boldsymbol{Q}_{\text{pot}}^{[s]}$, hyperparameter $\boldsymbol{\Lambda}_{0, \text{pot}}^{[s]}$ and design matrix $\boldsymbol{Z}_{\text{pot}}^{[s]}$. Gibbs samples in the following are then drawn from the distributions with respect to these potential structures.
	\ENDIF
	\FOR{MCMC samplings $t = 1$ to $T$}
		\STATE Take the estimates of iteration $s-1$ as the starting values.
		\STATE Sample $\hat{\boldsymbol{\gamma}}^{(t)}$ from $N(\boldsymbol{\mu}_{\gamma}, \boldsymbol{\Sigma}_{\gamma})$.
		\STATE Sample $\hat{\sigma}^{2(t)}$ from $IG(\tilde{a}, \tilde{b})$.
		\STATE Sample $\hat{\boldsymbol{Q}}^{(t)}$ from $IW(v, \boldsymbol{\Lambda})$.
	\ENDFOR
	\STATE Take the elementwise posterior modes of $\hat{\boldsymbol{\gamma}}^{(t)}$, $\hat{\sigma}^{2(t)}$ and $\hat{\boldsymbol{Q}}^{(t)}$ for all $T$ as the estimates $\hat{\boldsymbol{\gamma}}^{[s]}$, $\hat{\sigma}^{2[s]}$ and $\hat{\boldsymbol{Q}}^{[s]}$, respectively.
	\IF{$\text{MSE}_{k^*, \text{fixed}} < \text{MSE}_{k^*}$}
		\STATE Regard $X_{k^*}$ as an informative random effect, accept the potential structures of parameters as the new, and keep the estimates.
	\ELSE
		\STATE Reject the potential structures of parameters and reset them to the state of last boosting iteration. Subset the estimates by excluding $X_{k^*}$ relevant values and keep the rest.
	\ENDIF
	\STATE Update the predictor $\hat{\boldsymbol{\eta}}^{[s]} = \hat{\boldsymbol{\eta}}^{[s-1]} + \boldsymbol{X}_{k^*} \hat{\boldsymbol{\beta}}_{k^*}^{[s]} + \boldsymbol{Z}^{[s]} \hat{\boldsymbol{\gamma}}^{[s]}$
\ENDFOR
\RETURN $\hat{\boldsymbol{\eta}}^{[m_{\text{stop}}]}$
\end{algorithmic}
\end{algorithm}

\subsection{Model selection}
The conditional AIC (cAIC, \citet{greven2010}) is a good measure to perform model selection as briefly summarized in the introduction.
In boosting algorithms, the model at the end of each iteration can be regarded as a complete model, from which the cAIC can be calculated.
Thus we obtain a series of cAIC at the end of entire learning process and the best model should have the least cAIC.
However, possible extreme samples drawn from the conditional distributions in BayesBoost algorithm makes the global minimum of a cAIC series not reliable.
Even if the extremum is not considered, the randomness of the sampling process determines naturally cAICs to be a series with local waves and oscillations rather than a smooth curve.
We thus construct a special cAIC-based model selection criteria for BayesBoost.

To deal with the outliers of the cAIC series, we use the Hampel-filter \citep{pearson1999} that detects and removes outliers by using the Hampel identifier, which is a variation of the three-sigma rule of statistics but robust against outliers.
Specifically, the outliers are detected by the median absolute deviation in a moving-window, and then replaced with the local median.

The Hampel-filter is not a smoothing filter, hence local waves and oscillations will not disappear even though the filter is applied to the series.
Theoretically, all smoothing methods in the field of signals or time series can be applied to the cAIC series, decisions on the model choice can then be made based on the smoothed values.
Increasing the number of Gibbs samples at the cost of computing time can make the oscillations smaller, but cannot guarantee a disappearance.

Our solution of the problem is inspired by the early stopping meta-algorithm \citep{goodfellow2016} from deep learning, and we introduce a simple and non-smoothing method to find the local minimum from the oscillated cAIC series.
Due to the stepwise updating mechanism of boosting algorithms, there should exist little difference between the estimates at the stopping and its nearby iterations.
Therefore, early stopping can be understood as finding the stabilized or converged region first, and finding the smallest possible iteration from the region the next.
From the region perspective, the effect of oscillations caused by the randomness of sampling plays no important role anymore.
To control the scope of the region, we introduce a ``patience" hyperparameter $\alpha$ controlling the number of times to observe better iterations before stopping.
For example, given the cAIC of the model at iteration $s$ and patience $\alpha$, decision on whether $s$ should be the stopping iteration is based on if there exists smaller cAIC in the following iterations up to $s + \alpha$.
The existence of smaller value means a model improvement in the following iterations, i.e.\ the stabilized region has not been reached.
Otherwise the model has already been in good state and further updates may induce overfitting.

On the one hand, the patience parameter avoids immediately stopping if the right next cAIC larger than the current, which may simply cased by randomness in the samples.
It observes patiently the following iterations to check whether the increase of cAIC is random or a trend.
If it is a trend, then it stops, otherwise there must exists smaller values in the following iterations than the current.
On the other hand, if enough boosting iterations are performed, due to the randomness it can hardly avoid the case of the global minimum located at the overfitting area.
The ``patience" method seeks, however, the local minimum instead of the global, and the stopping iteration is usually located at the beginning of the stabilized region.

Though the burn-in is not necessary in BayesBoost as discussed above, its effect still exists, i.e.\ there exists big biases between the estimates and reality, because the stationary distribution has not reached within an acceptable error in this period.
Consequently, the cAIC series at early iterations may fluctuate drastically and sometimes affect the effectiveness of the patience parameter.
To avoid the effect of burn-in samples, we can force the model selection process starting from a relatively stable iteration.
There is no need to find a precise location, as long as the variation of cAICs is not obviously large, a rough number depending on the model complexity is already enough.

We present the special model selection method in Algorithm \ref{alg: stop}.

\begin{algorithm}
\caption{cAIC-based model selection criteria}
\label{alg: stop}
\begin{algorithmic}[1]
\STATE Apply Hampel-filter to the cAIC series.
\STATE Let $\alpha$ be the ``patience", the number of times to observe further cAICs before giving up.
\STATE Initialize from which iteration to start the selection $\zeta$.
\STATE Initialize $j = 0$, $i = \alpha + \zeta$ and $v = \infty$.
\STATE Initialize stopping iterations $s = i$.
\WHILE{$j < \alpha$}
	\STATE $i = i + 1$
	\IF{$\text{cAIC}_i < v$}
		\STATE $j = 0$
		\STATE $s = i$
		\STATE $v = \text{cAIC}_i$
	\ELSE
		\STATE $j = j + 1$
	\ENDIF
\ENDWHILE
\RETURN $s$.
\end{algorithmic}
\end{algorithm}
\section{Simulation}
\label{sec: sim}
In the following, two simulation studies are shown to demonstrate the performance of BayesBoost.
The first one compares the estimation accuracy for both random intercept and random slope model between BayesBoost and the enhanced gradient boosting algorithm \texttt{grbLMM} proposed by \citet{colin2021}.
The second one highlights the random effects selection of BayesBoost and explores its uncertainty estimation feature with a simulation example.

\subsection{Estimation accuracy}
The \texttt{grbLMM} algorithm shows the latest research results in applying gradient boosting technique to linear mixed models, but the covariance structure of the random effects must be specified in advance due to lacking an option for random effects selection.
The BayesBoost algorithm can mimick this behaviour by preserving the space of limiting the maximal number of random effects the final model contains or just giving a pre-defined covariance structure.
These can be achieved by closing the step \ref{alg: sketch5} in Algorithm \ref{alg: sketch} if the maximal number of random effects exceeds, or simply replacing the step with a pre-defined covariance structure (and skip the random effects selection, i.e.\ comparison of MSEs), and the latter is what we have done in this simulation to conduct a fair comparison.

We use the same setup as \citet{colin2021}, i.e., for individuals $i = 1, \cdots, 50$ and their replicates $j = 1, \cdots, 10$ and thus the total observations $n = 500$, the response in the random intercept model is drawn from
\begin{align*}
\boldsymbol{y}_i = 1 + 2\boldsymbol{x}_{i1} + 4\boldsymbol{x}_{i2} + 3\boldsymbol{x}_{i3} + 5\boldsymbol{x}_{i4} + \boldsymbol{\gamma}_{i0} + \boldsymbol{\varepsilon}_i,
\end{align*}
with $\boldsymbol{\gamma}_0 \sim N(0, \tau^2)$ , and in the random slope model is drawn from
\begin{align}
\boldsymbol{y}_i = 1 + 2\boldsymbol{x}_{i1} + 4\boldsymbol{x}_{i2} + 3\boldsymbol{x}_{i3} + 5\boldsymbol{x}_{i4} + \boldsymbol{\gamma}_{i0} + \boldsymbol{\gamma}_{i1} \boldsymbol{x}_{i3} + \boldsymbol{\gamma}_{i2} \boldsymbol{x}_{i4} + \boldsymbol{\varepsilon}_i,
\label{eq: sim1}
\end{align}
with $(\boldsymbol{\gamma}_{i0}, \boldsymbol{\gamma}_{i1}, \boldsymbol{\gamma}_{i2}) \sim N(\boldsymbol{0}, \boldsymbol{Q})$ where
\begin{align*}
\boldsymbol{Q} = \begin{pmatrix}
\tau^2	&	\tau^*	&	\tau^* \\
\tau^*	&	\tau^2	&	\tau^* \\
\tau^*	&	\tau^*	&	\tau^2
\end{pmatrix}.
\end{align*}
The simulations are evaluated for five different cases $p \in \{10, 25, 50, 100, 500\}$ ranging from low to high dimensions.
In both models, the random variables $\boldsymbol{x}_i, i = 1, \dots, p$ are standard normal distributed and only the first four variables are informative, specifically, $\boldsymbol{x}_1$ and $\boldsymbol{x}_2$ are cluster-constant covariates.
Moreover, $\boldsymbol{\varepsilon} \sim N(0, \sigma^2)$ with $\sigma = 0.4$ and $\tau \in \{0.4, 0.8, 1.6\}$. 
In the random slope model $\tau^*$ is chosen so that $\text{cor}(\boldsymbol{\gamma}_{ic}, \boldsymbol{\gamma}_{id}) = 0.6$ for all $c, d = 1, 2, 3$ holds.

The estimation accuracy is evaluated by the mean squared errors
\begin{align*}
\text{MSE}_{\boldsymbol{\theta}} := \|\boldsymbol{\theta} - \hat{\boldsymbol{\theta}}\|^2 \quad \text{and} \quad \text{MSE}_{\sigma^2} := \left(\sigma^2 - \hat{\sigma}^2\right)^2
\end{align*}
with $\boldsymbol{\theta} \in \{\boldsymbol{\beta}, \boldsymbol{\gamma}\}$.
For random intercept model, the variance $\tau^2$ is evaluated with $\text{MSE}_{\tau^2} := \left(\tau^2 - \hat{\tau}^2\right)^2$, and in case of the random slope model, the covariance matrix $\boldsymbol{Q}$ is measured by $\text{MSE}_{\boldsymbol{Q}} = \|\boldsymbol{Q} - \hat{\boldsymbol{Q}}\|_F$ where $\|\cdot\|$ denotes the Frobenius norm.
Performance of variable selection is evaluated by calculating the false positive rate (FP).
False negatives do not occur in both methods and hence are omitted.

Table \ref{tbl: ranInt} and \ref{tbl: ranSlo} summarizes the performance of 100 simulation runs for random intercept and random slope model respectively.
Estimates of \texttt{grbLMM} are taken at the stopping iterations determined by 10-folds cross-validation and the step-length $\nu = 0.1$ is set to each simulation runs.
BayesBoost is not an efficient algorithm due to the MCMC simulation and the calculation of cAIC. 
To increase the computing speed, we thus set a smaller MCMC sample $T = 30$ and a relatively larger step-length $\nu = 0.3$.
In addition, the stopping iteration in BayesBoost is determined by the cAIC-based selection criteria with the patience hyperparameter $\alpha = 3$, and the outliers are detected by the Hampel identifier with two-sigma rule instead of the common three.

Generally, \texttt{grbLMM} slightly outperforms BayesBoost in all mean squared error measures for both random intercept and random slope models, but their differences are not so noticeable. 
This indicates the BayesBoost's estimation accuracy is at the same level as \texttt{grbLMM}.
However, the false positive rates of \texttt{grbLMM} are obviously worse than BayesBoost in almost all cases.
This is reasonable because a model with more false positives usually overfits and therefore has better estimation accuracy.
The lower false positive rate in BayesBoost is mainly the consequence of the stopping criteria.
Changing the patience hyperparameter $\alpha$ to a higher value may lead to more accurate results at the cost of more false positives, but we can still expect a lower false positives from BayesBoost, because as stated above, the cAIC-based stopping criteria tends to select the beginning iterations at the stabilized region of cAIC series.
According to our research, if we select the global minimum of cAICs as the stopping iteration, we will get the false positive rates of BayesBoost similar to \texttt{grbLMM}, and consequently better MSEs than the cAIC-based results in the table.
This also demonstrates that BayesBoost will not perform worse than \texttt{grbLMM}.

\begin{table}
\caption{Results of model evaluation between \texttt{grbLMM} and BayesBoost in the random intercepts setup.}
\label{tbl: ranInt}
\centering
\begin{tabular}{ll|lllll|lllll}
\hline
\multirow{2}{*}{$\tau$}	&	\multirow{2}{*}{$p$}	&	\multicolumn{5}{c|}{\texttt{grbLMM}}	&	\multicolumn{5}{c}{BayesBoost}\\
&	&	$\text{MSE}_{\boldsymbol{\beta}}$	&	$\text{MSE}_{\boldsymbol{\tau}^2}$	&	$\text{MSE}_{\sigma^2}$	&	$\text{MSE}_{\boldsymbol{\gamma}}$	&	FP	&	$\text{MSE}_{\boldsymbol{\beta}}$	&	$\text{MSE}_{\boldsymbol{\tau}^2}$	&	$\text{MSE}_{\sigma^2}$	&	$\text{MSE}_{\boldsymbol{\gamma}}$	&	FP\\
\hline
\multirow{5}{*}{0.4}	&	10	&	0.013	&	0.001	&	$<$.001	&	1.192	&	0.48	&	0.015	&	0.002	&	$<$.001	&	1.314	&	0.15\\
	&	25	&	0.014	&	0.001	&	$<$.001	&	1.201	&	0.31	&	0.016	&	0.002	&	$<$.001	&	1.312	&	0.11\\
	&	50	&	0.015	&	0.001	&	0.001	&	1.194	&	0.20	&	0.017	&	0.002	&	$<$.001	&	1.302	&	0.08\\
	&	100	&	0.019	&	0.001	&	0.001	&	1.278	&	0.14	&	0.020	&	0.002	&	$<$.001	&	1.390	&	0.07\\
	&	500	&	0.021	&	0.001	&	0.001	&	1.241	&	0.04	&	0.022	&	0.001	&	0.001	&	1.430	&	0.04\\
\hline
\multirow{5}{*}{0.8}	&	10	&	0.043	&	0.014	&	$<$.001	&	2.570	&	0.49	&	0.041	&	0.017	&	$<$.001	&	2.993	&	0.17\\
	&	25	&	0.043	&	0.014	&	$<$.001	&	2.528	&	0.35	&	0.042	&	0.017	&	$<$.001	&	2.977	&	0.12\\
	&	50	&	0.050	&	0.012	&	0.001	&	2.759	&	0.24	&	0.049	&	0.017	&	$<$.001	&	3.227	&	0.09\\
	&	100	&	0.050	&	0.015	&	0.001	&	2.701	&	0.16	&	0.049	&	0.018	&	$<$.001	&	3.116	&	0.07\\
	&	500	&	0.057	&	0.015	&	0.001	&	2.837	&	0.05	&	0.067	&	0.022	&	0.003	&	3.240	&	0.05\\
\hline
\multirow{5}{*}{1.6}	&	10	&	0.155	&	0.230	&	$<$.001	&	7.850	&	0.47	&	0.152	&	0.238	&	$<$.001	&	9.493	&	0.17\\
	&	25	&	0.178	&	0.195	&	$<$.001	&	8.710	&	0.34	&	0.176	&	0.269	&	$<$.001	&	10.366	&	0.12\\
	&	50	&	0.176	&	0.259	&	0.001	&	8.413	&	0.29	&	0.174	&	0.297	&	$<$.001	&	9.787	&	0.09\\
	&	100	&	0.174	&	0.238	&	$<$.001	&	8.416	&	0.14	&	0.172	&	0.310	&	$<$.001	&	9.835		&	0.06\\
	&	500	&	0.166	&	0.255	&	0.001	&	7.823	&	0.05	&	0.165	&	0.277	&	0.001	&	9.162	&	0.04\\
\hline
\end{tabular}
\end{table}

\begin{table}
\caption{Results of model evaluation between \texttt{grbLMM} and BayesBoost in the random slope setup.}
\label{tbl: ranSlo}
\centering
\begin{tabular}{ll|lllll|lllll}
\hline
\multirow{2}{*}{$\tau$}	&	\multirow{2}{*}{$p$}	&	\multicolumn{5}{c|}{\texttt{grbLMM}}	&	\multicolumn{5}{c}{BayesBoost}\\
&	&	$\text{MSE}_{\boldsymbol{\beta}}$	&	$\text{MSE}_{\boldsymbol{Q}}$	&	$\text{MSE}_{\sigma^2}$	&	$\text{MSE}_{\boldsymbol{\gamma}}$	&	FP	&	$\text{MSE}_{\boldsymbol{\beta}}$	&	$\text{MSE}_{\boldsymbol{Q}}$	&	$\text{MSE}_{\sigma^2}$	&	$\text{MSE}_{\boldsymbol{\gamma}}$	&	FP\\
\hline
\multirow{5}{*}{0.4}	&	10	&	0.020	&	0.013	&	0.002	&	4.482	&	0.46	&	0.050	&	0.014	&	0.004	&	4.434	&	0.04\\
	&	25	&	0.022	&	0.012	&	0.003	&	4.530	&	0.27	&	0.052	&	0.014	&	0.004	&	4.493	&	0.02\\
	&	50	&	0.023	&	0.012	&	0.003	&	4.551	&	0.18	&	0.051	&	0.014	&	0.004	&	4.500	&	0.02\\
	&	100	&	0.025	&	0.012	&	0.003	&	4.536	&	0.10	&	0.057	&	0.014	&	0.004	&	4.557	&	0.02\\
	&	500	&	0.027	&	0.011	&	0.003	&	4.453	&	0.03	&	0.050	&	0.010	&	0.003	&	4.293	&	0.01\\
\hline
\multirow{5}{*}{0.8}	&	10	&	0.072	&	0.124	&	0.002	&	6.923	&	0.44	&	0.124	&	0.140	&	0.036	&	7.983	&	0.17\\
	&	25	&	0.073	&	0.121	&	0.003	&	7.015	&	0.28	&	0.137	&	0.143	&	0.048	&	8.314	&	0.12\\
	&	50	&	0.074	&	0.119	&	0.003	&	6.956	&	0.17	&	0.143	&	0.143	&	0.044	&	8.385	&	0.08\\
	&	100	&	0.078	&	0.094	&	0.003	&	7.060	&	0.11	&	0.149	&	0.115	&	0.046	&	8.530	&	0.06\\
	&	500	&	0.082	&	0.124	&	0.003	&	6.953	&	0.04	&	0.184	&	0.136	&	0.076	&	8.897	&	0.04\\
\hline
\multirow{5}{*}{1.6}	&	10	&	0.280	&	1.829	&	0.002	&	16.970	&	0.41	&	0.299	&	2.145	&	0.008	&	17.109	&	0.18\\
	&	25	&	0.277	&	1.808	&	0.002	&	16.605	&	0.29	&	0.299	&	2.139	&	0.008	&	17.169	&	0.13\\
	&	50	&	0.294	&	1.435	&	0.002	&	17.124	&	0.19	&	0.330	&	1.741	&	0.012	&	18.271	&	0.09\\
	&	100	&	0.299	&	1.852	&	0.003	&	16.682	&	0.14	&	0.341	&	2.012	&	0.041	&	17.281	&	0.06\\
	&	500	&	0.320	&	1.804	&	0.003	&	17.658	&	0.04	&	0.341	&	1.909	&	0.014	&	17.578	&	0.04\\
\hline
\end{tabular}
\end{table}

\subsection{Random effects selection}
To explore the performance of random effects selection in BayesBoost as well as other features, we use the same simulation settings as in the equation \eqref{eq: sim1} but select only one typical setup with $\tau = 0.8$ and $p = 50$, and let all covariates (except for the cluster-constant covariates) be the participants for the random effects.
In other words, we do not need to specify the covariance structure of random effects in advance, but let the algorithm choose automatically.

Table \ref{tbl: mod_eval} shows the model evaluation measures as above under three stopping conditions, the first two use the cAIC-based stopping criteria with patience hyperparameter $\alpha = 3$ and $5$, and the third uses the global minimum of the cAIC series.
Note that the random effects $\boldsymbol{\gamma}$, its covariance matrix $\boldsymbol{Q}$ and false positive/negative rates are evaluated between the truth and the corresponding subset of estimates, as the estimate may have a larger or smaller matrix structure due to the random effects selection.

Due to the same setup, we can compare the results in Table \ref{tbl: mod_eval} and the corresponding row ($\tau = 0.8$ and $p = 50$) in Table \ref{tbl: ranSlo}.
For $\alpha = 3$, which is also the setup in Table \ref{tbl: ranSlo}, the MSEs in the automatic random effects selection setting are slightly worse than in the pre-defined setting, but the former has better false positive rates of fixed effects.
For a more conservative stopping hyperparameter $\alpha = 5$, the false positive rates of fixed effects in Table \ref{tbl: mod_eval} are on the same level as in Table \ref{tbl: ranSlo}, but the MSEs of the former are even better than the latter.
In terms of the usage of global minimum, the much better MSEs are gained at the cost of lots of false positives of fixed effects.
In general, the increase or decrease of MSEs and false positives behaves in a predictable manner, i.e. the cAIC-based methods tend to stop early before reaching the global minimum, early stopped results tend to be underfitted and the results with more false positives have lower MSEs.
Therefore, from this comparison we can find that even the random effects are not given in advance, the BayesBoost algorithm can still capture the structures effectively and give competitive outcomes.


Table \ref{tbl: mod_eval} lists also the false negative rates of random effects $\boldsymbol{\gamma}$.
The same low value of 0.010 in all three circumstances indicates that neither or one of the two informative random effects is selected only in a few simulation runs, and in most cases, they are successfully chosen by the algorithm.
This shows also that BayesBoost is effective in selecting random effects.
As there is no occurrence of false positives of $\boldsymbol{\gamma}$ and false negatives of $\boldsymbol{\beta}$, they are hence omitted from the table.
But we still have to emphasize that the assumption that only fixed effects can have random effects ensures that the false positives of $\boldsymbol{\gamma}$ will not exceed that of $\boldsymbol{\beta}$.



\begin{table}
\caption{Model evaluations for $\tau = 0.8$ and $p = 50$ for different stopping methods. The first two rows show the results at the stopping iteration determined by the cAIC-base stopping criteria with patience hyperparameter $\alpha = 3$ and $5$. The last row shows the results of model with the least cAIC.}
\label{tbl: mod_eval}
\centering
\begin{tabular}{l|llllll}
\hline
	&	$\text{MSE}_{\boldsymbol{\beta}}$	&	$\text{MSE}_{\boldsymbol{Q}}$	&	$\text{MSE}_{\sigma^2}$	&	$\text{MSE}_{\boldsymbol{\gamma}}$	&	$\text{FP}_{\boldsymbol{\beta}}$	&	$\text{FN}_{\boldsymbol{\gamma}}$\\
\hline
$\alpha = 3$	&	0.185	&	0.134	&	0.080	&	8.559	&	0.043	&	0.010\\
$\alpha = 5$	&	0.100	&	0.129	&	0.018	&	7.278	&	0.085	&	0.010\\
Min.			&	0.077	&	0.137	&	0.004	&	6.913	&	0.333	&	0.010\\
\hline
\end{tabular}
\end{table}

To explore the characteristics of BayesBoost, we take a typical simulation from the 100 runs as an example, and the stopping iteration discussed following is determined by the cAIC-based method with $\alpha= 5$.
In this example, except for all the informative fixed and random effects, the model at the stopping iteration contains additionally 6 non-informative fixed effects.

\begin{figure}
\centering
\includegraphics[width=\textwidth]{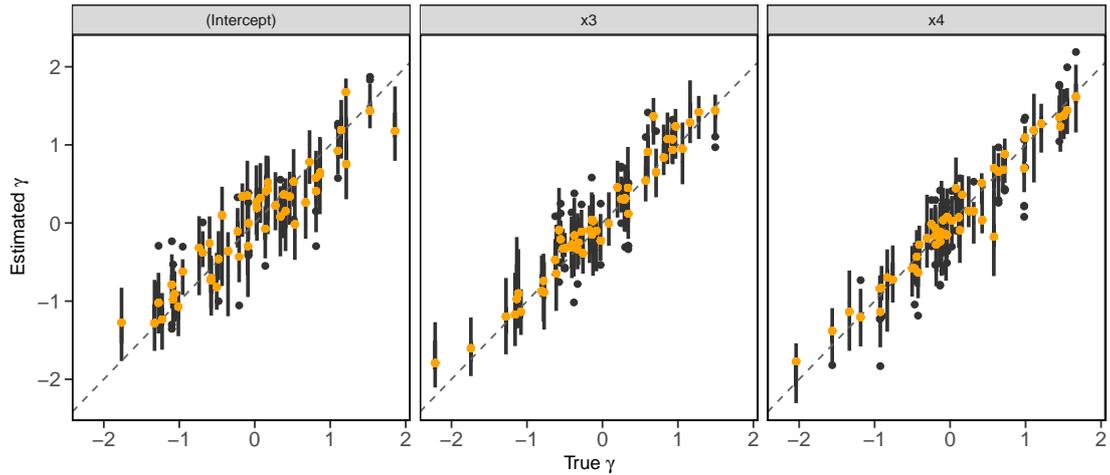}
\caption{Distributions of the Gibbs samples for all individuals at the stopping iteration. Black lines summarize the distribution of the samples with outliers marked with the black dot. Orange points denotes the sample modes (posterior modes) of each individual. The reference dashed grey line marks the location where the estimates coincides with the truth.}
\label{fig: gamma_dist}
\end{figure}

Figure \ref{fig: gamma_dist} illustrates the distribution of random effects at the stopping iteration.
According to the algorithm setup, $T = 30$ MCMC samples for each individual are summarized by a black line (and outliers marked with black points).
The orange points represents the (posterior) modes of the samples and serves as the estimates of random effects $\hat{\boldsymbol{\gamma}}$.
The theoretical ideal estimation is given through the dashed grey reference line.

The fact that the estimated $\hat{\boldsymbol{\gamma}}$ matches the true $\boldsymbol{\gamma}$ quiet well shows that the BayesBoost is effective in the random effect estimation first of all.
But the more important outcomes brought by BayesBoost is the uncertainty estimation of random effects measured by MCMC samples.
Given the sample distribution, parameter estimation using boosting technique is no more an estimation point alone, but we can further know to what extent to accept the estimates.
Hypothesis test, credible interval as well as other Bayesian statistics of interest for the random effects (and also for their covariance matrix $\boldsymbol{Q}$) can thus be established based on the samples.
Increasing the size of MCMC samples $T$ may produce a more accurate estimate, 
but even with a small sample size as in this example, the outcomes are still satisfactory, because the majority of the distribution lines cover the truth.
Permutation or bootstrap techniques may help the common boosting methods to get access to the uncertainties of estimates, but they suffer from the bias induced by shrinkage.
BayesBoost, however, gets rid of the shrinking estimation method but estimates random effects all at once in each iteration.
This makes the uncertainties obtained from BayesBoost more reliable.

\begin{figure}
\centering
\begin{subfigure}[b]{.48\textwidth}
\includegraphics[width=\textwidth]{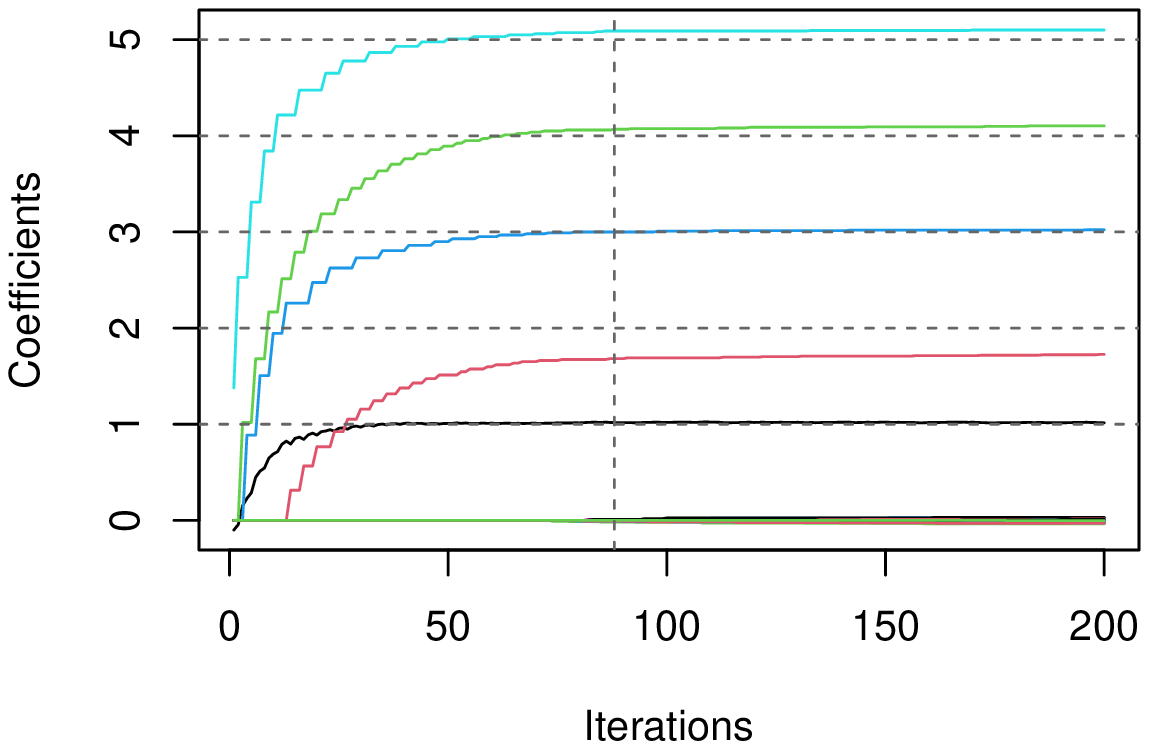}
\caption{Fixed effects $\hat{\boldsymbol{\beta}}$}
\label{fig: coef_beta}
\end{subfigure}
\begin{subfigure}[b]{.48\textwidth}
\includegraphics[width=\textwidth]{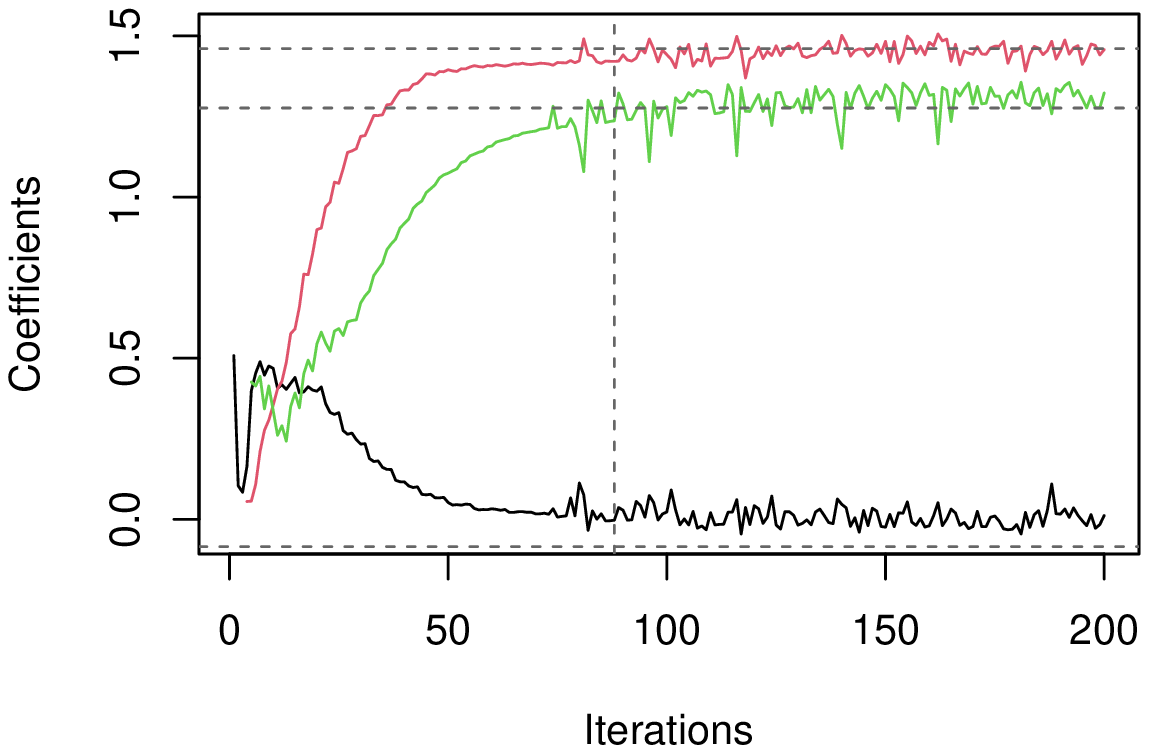}
\caption{Random effects $\hat{\boldsymbol{\gamma}}_i$ of an individual $i$}
\label{fig: coef_gamma}
\end{subfigure}
\begin{subfigure}[b]{.48\textwidth}
\includegraphics[width=\textwidth]{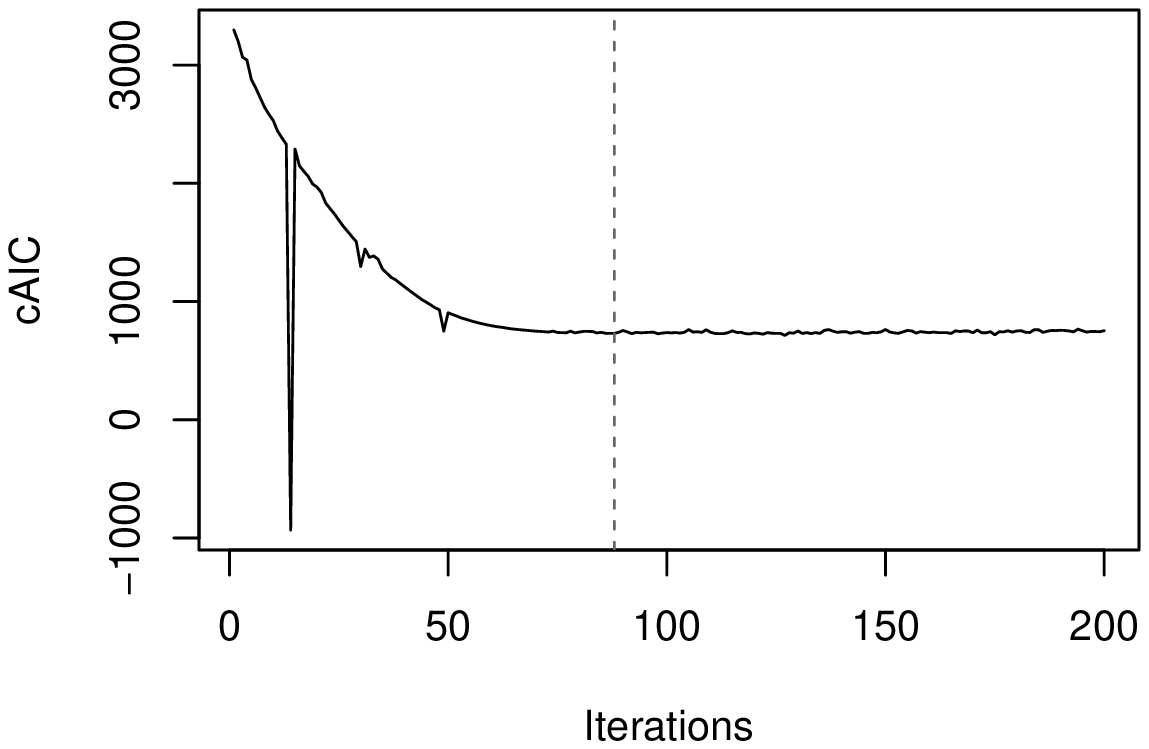}
\caption{cAIC}
\label{fig: cAIC}
\end{subfigure}
\begin{subfigure}[b]{.48\textwidth}
\includegraphics[width=\textwidth]{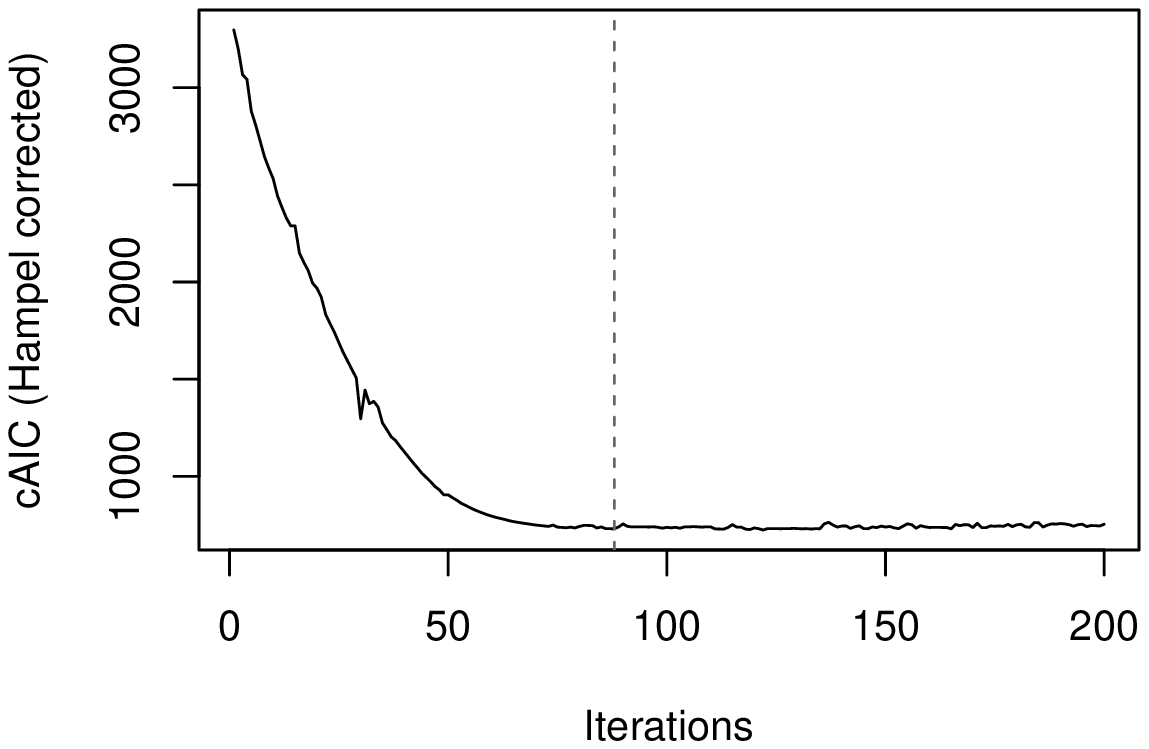}
\caption{cAIC (Hampel corrected)}
\label{fig: cAIC_hampel}
\end{subfigure}
\caption{The estimated coefficients of fixed and random terms in each boosting iteration with the stopping iteration marked with the dashed vertical line and the true values marked with the vertical dashed lines. Plot (\subref{fig: coef_beta}) shows the estimates of fixed effects $\hat{\boldsymbol{\beta}}$. Plot (\subref{fig: coef_gamma}) displays the estimated random effects $\hat{\boldsymbol{\gamma}}_i$ of an individual $i$, where each line in this plot is drawn by the the sample modes. The black curve in both plots represents the model intercept and random intercept respectively. The red and green curves in plot (\subref{fig: coef_gamma}) represent the estimates for random slope $\boldsymbol{x}_4$ and $\boldsymbol{x}_3$, while they represent the fixed effect $\boldsymbol{x}_1$ and $\boldsymbol{x}_2$ in plot (\subref{fig: coef_beta}), respectively. Plot (\subref{fig: cAIC}) and (\subref{fig: cAIC_hampel}) shows the cAICs and through Hampel-filter corrected cAICs of each models.}
\label{fig: coef}
\end{figure}

Figure \ref{fig: coef} shows the estimated coefficient paths of each covariates along with the boosting iterations for fixed effects $\hat{\boldsymbol{\beta}}$ and the random effects $\hat{\boldsymbol{\gamma}}_i$ (posterior mode) of an individual $i$.
The stopping iteration is marked with the vertical black dashed line.
As discussed above, BayesBoost estimates random effects not through the sum up of learning pieces but altogether as a whole.
So the shrinkage of coefficients can only be interpreted from the estimation of $\boldsymbol{\beta}$ but not from the random effects.
The convergence behaviour of random effects is actually not the consequence of shrinkage.
The other important boosting feature, variable selection, is shared by both fixed and random effects.
In this example, up to the stopping iteration, all informative fixed and random effects as well as 6 of the noise fixed effects are included into the final model with very small coefficients.

If we take a closer look at Figure \ref{fig: coef}(\subref{fig: coef_gamma}), we will find that the coefficient paths can be roughly divided into three periods.
The first period from the beginning lasts to about 20 iteration.
The estimates in this period oscillate heavily mainly due to the burn-in of MCMC simulations.
The second period lasts up to 88 iteration, which is also the stopping iteration, or a little earlier.
In this period, along with the convergence of fixed effects, the random effects converges relatively smoothly to the truth.
The remaining iterations can be grouped as the third period, as estimation for the random effects in this period is no more stable, but waves around their converged values. 
This is mainly caused by occurrence of more and more noise fixed effects.
Specifically, the first noise fixed effects comes at the 74 iteration, which is earlier than the stopping iteration but the exactly beginning of oscillation in this period.

\begin{figure}
\centering
\includegraphics[width=.6\textwidth]{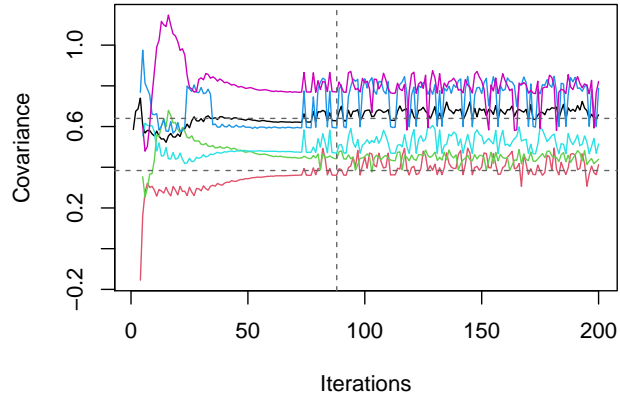}
\caption{Estimates of covariance $\boldsymbol{Q}$ in each iteration. According to the model specification, the true variance (diagonal of $\boldsymbol{Q}$) for each random effects is 0.64, the true covariance between the random effects (off-diagonal) is all 0.384, and both reference values are marked with the dashed grey horizontal line. The dashed vertical line indicates the stopping iteration. At the stopping iteration, the above three lines are the variance of random effects, where the black line is the variance of random intercept. The under three lines are the covariance between random effects.}
\label{fig: Q}
\end{figure}

As mentioned above that the seemingly ``convergence" of random effects in the second period cannot be interpreted as the shrinking estimation as fixed effects.
Since the random effects in each BayesBoost iteration capture the residual information as much as possible, each model during this period is already a mature model.
This can be observed more clearly from the similar graphic for the covariance matrix $\boldsymbol{Q}$ in Figure \ref{fig: Q}.
The covariance between random effects changing little from about the 40 iteration to the stopping iteration indicates the covariance matrix have already been in good state.
However, the coefficients of random effects in Figure \ref{fig: coef}(\subref{fig: coef_gamma}) still shows the convergence behavior during this period.
This phenomenon is actually the consequence of the changes of fixed effects in this period.
As discussed above, random effects should only explain the response that cannot be explained by the fixed effects.
Based on this statement the estimation for the random effects gives in to that for the fixed effects.
Practically, as fixed effects converge, information extracted by the random effects in the previous iteration will be spitted out in the later iterations and forms the seemingly convergence behavior.

Figure \ref{fig: coef}(\subref{fig: cAIC}) and (\subref{fig: cAIC_hampel}) shows also the cAIC and through Hampel-filter corrected cAIC of model in each boosting iteration.
The randomness of the estimation procedure can be observed clearly from these plots.
On the one hand, there is an obvious outlier in plot (\subref{fig: cAIC}) at iteration 14, and it is successfully detected and replaced by the Hampel-filter.
On the other hand, even the Hampel-filter is applied, we can still observe small waves and oscillations in the series.
That explains why it is necessary to establish a stopping criteria based on the random cAICs, because we cannot guarantee whether a global minimum is true or random.

In addition, Figure \ref{fig: Q} is also drawn by the elementwise posterior modes of the covariance samples.
That means the Bayesian analysis for the random effects applies also to their covariance structure.

\section{Application}
\label{sec: app}
To test the application of the algorithm, we used it on a neuroscientific data set measuring sound perception with magnetoencephalography (MEG).
Data from neuroscience are especially interesting because of their multidimensionality, and because they contain a multitude of (possible) measures on usually relatively few participants or observations - which is a problem for neuroscientists looking for valid statistical methods that adequately deal with extensive multiple comparisons or, even better, are able to focus the shear endless amount of possible measures to a few relevant ones.
We show here that BayesBoost offers the possibility for neuroscientists to meaningfully detect and select relevant brain variables in their experiments without prior choices made from possibly insufficient theory.

The dataset stems from a cognitive neuroscience experiment, where 20 healthy subjects heard, among some visual stimuli, four different annoying, tinnitus-like sounds.
Each sound was presented 16 times, yielding 64 trails in total per subject.
After each trial or presentation, subjects provided a subjective rating of the loudness of the sound (along with 3 other rating questions, which are not included here).
Ratings were input on a visual analog scale, allowing continuous ratings from 0 to 1.
Additionally, subjects filled questionnaires on psychological measures, like emotion regulation.

The goal here was to use BayesBoost to find meaningful variables among a myriad of oscillatory power measures in the brain, that could explain, together with psychological and experimental variables in the model, the loudness perception of the tones as reflected in their ratings.

Oscillatory power in different frequency bands is one type of measure reflecting brain activity.
A typical pipeline for obtaining power measures from MEG data is described in \citet{van2018analysis}.
Usually, scientist pre-select a band or regions of interest, and try to narrow down a relevant time point (activation peak) during the trial.
To test BayesBoost's ability to do this for the investigator, and crucially to find meaningful variables algorithmically, no pre-selections were made beyond calculating aggregate oscillatory power values for 4 time segments of 2 seconds duration during each tone trial (total length 8 seconds), in 80 different brain regions (comprising the Desikan-Killany parcellation \citep{desikan2006automated} with 34 regions, and 6 additional subcortical regions per hemisphere (left and right)), and for 6 different frequency bands (from theta to high gamma).
The power measures were converted into $\text{fT}^2$ and log base 10 transformed to make them linear.

Covariates included in the data set were:
\begin{itemize}
\item Subject ID (cluster id)
\item Gender - cluster-constant variable
\item Emotion regulation score (ER\_ges) - cluster-constant variable
\item Tone (Ton) - experimental variable (which of 4 tones was played)
\item 1920 brain variables reflecting oscillatory power at different time points during the trail ($n=4$) [Ton\_Part1,$\cdots$, Ton\_Part4], in different brain areas ($n=80$; 40 each in the left and right hemisphere), and in different frequency bands ($n=6$) [theta, alpha, beta\_low, beta\_high, gamma, gamma\_high]
\end{itemize}

As only trials with complete data were used (missing data excluded), there were $N=1161$ total observations (for $n=20$ clusters) in the dataset, and 1924 variables to enter the algorithm.
To take a balance between the learning speed and the estimation accuracy, we take the step-length $\nu = 0.3$ and draw $T = 300$ MCMC samples in each boosting iteration.

\begin{table}
\centering
\caption{Selected variables (grouped by the tone parts) and their coefficients (fixed effects).}
\label{tbl: ton}
\begin{tabular}{l|l|r}
\hline
&&	Coefficients	\\
\hline
&	(Intercept)	&	0.5500	\\
\hline
\multirow{2}{*}{Experiment}	&	ER\_ges	&	-0.0154	\\
&	Ton	&	-0.0133	\\
\hline
\multirow{6}{*}{Part 1}	&	frontalpole.lh\_beta\_high	&	-0.0112	\\
&	frontalpole.lh\_gamma	&	-0.0105	\\
&	lateralorbitofrontal.lh\_gamma	&	-0.0046	\\
&	superiorfrontal.lh\_alpha	&	-0.0038	\\
&	paracentral.lh\_alpha	&	-0.0073	\\
&	parahippocampal.lh\_beta\_low	&	0.0040	\\
\hline
\multirow{8}{*}{Part 2}	&	frontalpole.lh\_beta\_high	&	-0.0113	\\
&	lateralorbitofrontal.lh\_theta	&	-0.0058	\\
&	lateralorbitofrontal.rh\_alpha	&	-0.0051	\\
&	rostralanteriorcingulate.lh\_alpha	&	-0.0038	\\
&	caudalanteriorcingulate.lh\_alpha	&	-0.0039	\\
&	rostralmiddlefrontal.lh\_theta	&	-0.0060	\\
&	Putamen.lh\_gamma	&	0.0041	\\
&	cuneus.lh\_beta\_low	&	-0.0047	\\
\hline
\multirow{2}{*}{Part 3}	&	Caudate.rh\_alpha	&	-0.0050	\\
&	Amygdala.rh\_beta\_high	&	0.0048	\\
\hline
\multirow{3}{*}{Part 4}	&	precentral.lh\_beta\_low	&	0.0040	\\
&	temporalpole.lh\_theta	&	0.0059	\\
&	superiortemporal.lh\_alpha	&	0.0048	\\
\hline
\end{tabular}
\end{table}

The selected fixed effects are listed in Table \ref{tbl: ton}.
The BayesBoost algorithm produces a model that combines general effects of personal dispositions, experimental manipulation, and specific brain processes, while also suggesting individual processes in the neurophysiological processes involved in perceiving and judging the annoying tones:
The cluster-constant variable of subjects' emotion regulation score shows that people with better regulation skills rate tones as less loud.
There is also a general effect of specific tones being generally perceived as louder/less loud than others.
The final model does not contain the gender variable indicates the gender of the participants has no noticeable effect on the loudness perception of the tones.
The effects of oscillatory power in the brain then seem to reflect some relevant general processes, which, because they are spread over groups of similar/related variables, seem to indicate individual expressions of these general functional processes over subjects.
During the first two parts of the tone, a group of frontal and attention-related brain regions appear with negative coefficients, possibly reflecting regulation mechanisms to counter the annoyance of the tones.
They show relations of power increases in different frequency bands in e.g. orbitofrontal areas or the anterior cingulate to decreased tone loudness ratings.
Then, some auditory (e.g.\ superiortemporal) and memory- or emotion-related areas (e.g.\ parahippocampal and subcortical regions) appear with positive coefficients throughout the model, probably reflecting auditory and emotional processes related to the annoyance of the tones and, hence, an increased loudness perception.
Finally, in part 4 of the tones, we see an increase in motor cortex low beta power with increased loudness ratings, possibly reflecting the preparation of the rating via button press.

\begin{figure}
\includegraphics[width=\textwidth]{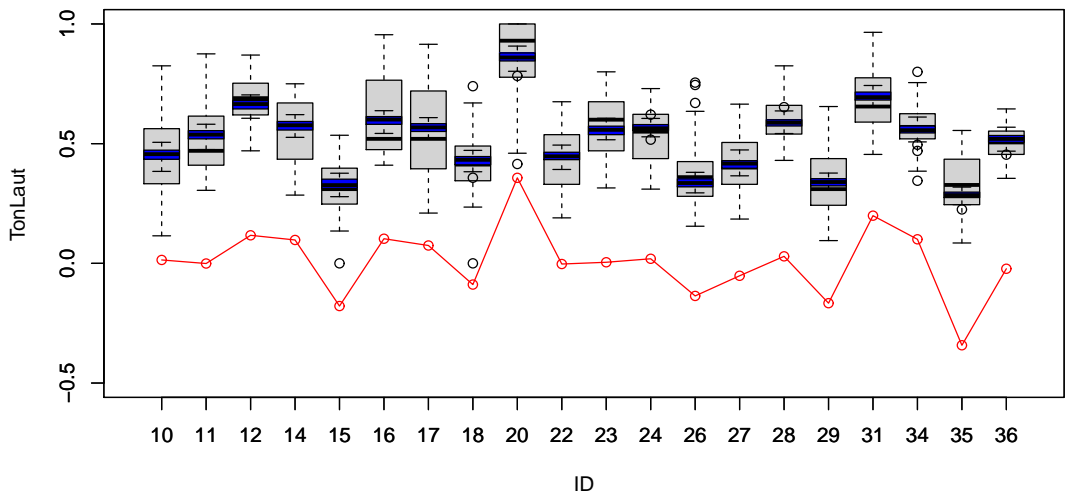}
\caption{Boxplot of the rating of loudness of sounds separated by each participant. The grey boxplots demonstrates the distribution of true values, and the blue ones illustrates the in-sample predictions. The red lines shows the random intercept of each participant estimated by BayesBoost.}
\label{fig: tone}
\end{figure}

Figure \ref{fig: tone} illustrates the distribution of the loudness ratings of the sounds separated by each participant.
The final model estimated by the BayesBoost algorithm contains only random intercepts, and they are highlighted by the red points and lines.
In general, though the ratings of loudness exhibit obviously individual differences, the differences are well detected and captured by the estimated random intercepts.

Overall, the variable selection of the BayesBoost algorithm seems to reflect a holistic picture of psychological, experimental, and brain measures in sound perception, while also displaying the variety of individually varying, but functionally similar brain processes involved in a task.
Especially this latter feature is something that could be fruitfully further explored in cooperations between statisticians and neuroscientists, as the architecture of BayesBoost with its possibility to model random effects and slopes in a fine-grained fashion makes it a promising candidate for more fine-tuned model-building with neuroscientific data.
\section{Discussion and Outlook}
\label{sec: summary}
One of the most important reasons for Boosting techniques being widely used in statistics is due to its appealingly direct and effective variable selection feature.
However, as a method originated from machine learning, it lacks straight forward ways to construct estimators for the precision of the parameters such as variance or confidence intervals like common statistical approaches.
Thanks to the development in computer science, the Bayesian inference has grown immensely in the last decades and rendered possible an extreme amount of new types of models.
But it very often fails to give precise and unambiguous guidelines for the selection of variables.

This paper proposes a new inference method, BayesBoost, by combining the two concepts to benefit from both worlds.
On the one hand, the new approach has totally kept the variable selection feature of model-based Boosting methods, so that the parameter estimation and both fixed and random effects selection can be performed simultaneously.
On the other hand, variation of random effects is accessible through the BayesBoost estimation, which is not possible in the conventional boosting framework.
The effectiveness of BayesBoost can be observed from the simulation and empirical studies. 
However, as a new attempt, it leaves also lots of open questions.

As a point on the application we want to mention the necessity to dig deeper into the field of neuroscientific data analysis with this type of model. 
We are sure that, by adjusting the method further to the specific problems posed by this kind of multidimensional data, we can find important patterns in MEG or other neurophysiological measures.

We further feel that there is direct need in the field of methods developement: Firstly, this paper only focuses on the linear mixed models, whether it is possible to extend the BayesBoost concept to other common statistical models is one of the most important questions.
In other words, a more generalized BayesBoost framework needs be proposed instead of the current linear mixed model specific version.

Secondly, even for the linear mixed models, BayesBoost only fills in the uncertainty estimation blank of the random effects part left by the Boosting framework, that of the fixed effects part is still unsolved.
It seems possible to address the problem by optimizing regression models with base-learners one by one by Bayesian inference, but still more work needs to be done.

Thirdly, although BayesBoost provides the interesting uncertainty estimates for the random effects, which makes hypothesis test and credible interval possible, it also leads to the uncertainties in the model evaluation, i.e.\ the randomness of cAIC, or more specifically, the randomness of the global minimum of cAIC.
This paper introduces a cAIC-based model selection criteria, which determines the stopping iteration based on the stabilized region.
But even if the filtering method is not used, the additional formulation of a patience hyperparameter is rather unsatisfactory.
A more mature stopping criteria based on the oscillatory cAICs or a method similar to cross-validation still requires further research.

Further studies should also include the performance of hypothesis test and credible interval.
Tests about the effectiveness of the approach on the non-linear or spatial base-learners are also meaningful.
Bayesian method has already the reputation of producing its results very slowly, but implementing Bayesian inference inside each Boosting iteration will make the situation more severe, therefore the algorithm optimization is also an important task.

\subsection*{Acknowledgements}
This work was supported by the Freigeist-Fellowships of Volkswagen Stiftung, project ``Bayesian Boosting - A new approach to data science, unifying two statistical philosophies". Boyao Zhang performed the present work in partial fulfilment of the requirements for obtaining the degree ``Dr. rer. biol. hum." at the Friedrich-Alexander-Universit\"at Erlangen-N\"urnberg (FAU).

\bibliographystyle{apalike} 
\bibliography{reference}

\appendix
\section{Appendix}
\subsection{BayesBoost algorithm}
\label{apx: bayesboost}
\begin{breakablealgorithm}
\caption{BayesBoost for LMM}
\label{alg: bayesboost}
\begin{algorithmic}[1]
\STATE Initialize $\hat{\boldsymbol{y}}_i^{[0]} = \frac{1}{n_i}\sum_{j = 1}^{n_i}(y_{ij})$, $i \in \{i, \cdots, m\}, j \in \{1, \cdots, n_i\}$.
\STATE Initialize $\hat{\beta}_0^{[0]} = \frac{1}{n} \sum_{i = 1}^m \sum_{j = 1}^{n_i} y_{ij}$, where $n = \sum_{i=1}^m n_i$.
\STATE Initialize $\boldsymbol{R}^{(0)} = (\sigma^2)^{(0)}\boldsymbol{I} = 1$.
\STATE Initialize $\boldsymbol{Q}^{[0]} = 1$
\STATE Initialize $\boldsymbol{\Lambda}_0^{[0]} = 1$, $a = b = 0.001$ and $v_0 = 1$.
\STATE Initialize a random effects set $E = \{\text{ranInt}\}$ that contains only the random intercept.
\STATE Construct a design matrix $\tilde{\boldsymbol{Z}}$ that involves only random intercepts. If there exists cluster-constant covariates, transform $\tilde{\boldsymbol{Z}}$ to the correction matrix $\boldsymbol{Z}$
	\begin{align*}
	\boldsymbol{Z}^{[0]} &= \tilde{\boldsymbol{Z}} - \boldsymbol{X}_c(\boldsymbol{X}_c^T \boldsymbol{X}_c)^{-1} \boldsymbol{X}_c^T \tilde{\boldsymbol{Z}},
	\end{align*}
so that $\boldsymbol{Z}^{[0]} \perp \boldsymbol{X}_c$, where $\boldsymbol{X}_c \subset \boldsymbol{X}$ denotes the cluster-constant covariates, otherwise $\boldsymbol{Z}^{[0]} = \tilde{\boldsymbol{Z}}$.
\FOR{Boosting iteration $s = 1$ to $m_{\text{stop}}$} 
	\STATE Compute the model residuals
		\begin{align*}
		\boldsymbol{u}^{[s]} = \boldsymbol{y} - \hat{\boldsymbol{y}}^{[s - 1]}.
		\end{align*}
	\STATE Fit every covariates $X_k, k \in \{1, \cdots, p\}$ separately to the residuals $\boldsymbol{u}^{[s]}$ with linear regression model to get their coefficients $\hat{\boldsymbol{\beta}}_{k}$ (with intercept).
	\STATE Select the best-fitting base-learner $X_{k^*}$ that results in the least residual sum of squares,
		\begin{align*}
		\text{MSE}_{k, \text{fixed}} &= \frac{1}{n} \sum_{i = 1}^n\left(u_i - \boldsymbol{X}_{ik}\hat{\boldsymbol{\beta}}_k\right)^2 \\
		k^* &= \argmin_{k \in \{1, \cdots, p\}} \text{MSE}_{k, \text{fixed}},
		\end{align*}
	where $\hat{\boldsymbol{\beta}}_k = (\hat{\beta}_0, \hat{\beta}_k)^T$ and $\boldsymbol{X}_k = (1, X_k)$.
	\STATE Update the fixed effects $\hat{\boldsymbol{\beta}}^{[s]}$ through $\hat{\boldsymbol{\beta}}_{k^*}$ with an appropriate step length $\nu$
		\begin{align*}
		\hat{\boldsymbol{\beta}}^{[s]} = \hat{\boldsymbol{\beta}}^{[s-1]} + \nu \hat{\boldsymbol{\beta}}_{k^*}.
		\end{align*}
	\IF{$X_{k^*}$ is not in the random effects set $E$}
		\STATE Treat $X_{k^*}$ as a potential random effect and construct a potential design matrix $\boldsymbol{Z}_{\text{pot}}^{[s]}$ by inserting a corrected matrix $\boldsymbol{Z}_{k^*}$ into $\boldsymbol{Z}^{[s-1]}$ with an appropriate order
			\begin{align*}
			\underbrace{\boldsymbol{Z}_{\text{pot}}^{[s]}}_{n \times (M+m)} = \underbrace{\boldsymbol{Z}^{[s - 1]}}_{n \times M} \cup \underbrace{\boldsymbol{Z}_{k^*}}_{n \times m},
			\end{align*}
		where $M$ is a multiple of $m$, and $M = m$ for $s = 1$. The corrected matrix obtained through
			\begin{align*}
			\boldsymbol{Z}_{k^*} = \tilde{\boldsymbol{Z}}_{k^*} - X_{k^*} \left(X_{k^*}^T X_{k^*}\right)^{-1} X_{k^*}^T \tilde{\boldsymbol{Z}}_{k^*}
			\end{align*}
		is orthogonal to $X_{k^*}$, and $\tilde{\boldsymbol{Z}}_{k^*}$ is the $(n \times m)$ design matrix of $X_{k^*}$.
		\STATE Construct a potential covariance matrix
			\begin{align*}
			\boldsymbol{Q}_{\text{pot}}^{[s]} = \text{diag}(\boldsymbol{Q}^{[s - 1]}, 1)
			\end{align*}
			with $1$ as the initialized variance of random effect $X_{k^*}$ and thus
			\begin{align*}
			\boldsymbol{G}^{(0)} = \text{blockdiag}(\boldsymbol{Q}_{\text{pot}}^{[s]}, \cdots, \boldsymbol{Q}_{\text{pot}}^{[s]}, \cdots, \boldsymbol{Q}_{\text{pot}}^{[s]}).
			\end{align*}
		\STATE Construct a potential matrix $\boldsymbol{\Lambda}_{0, \text{pot}}^{[s]} = \text{diag}(\boldsymbol{\Lambda}_0^{[s - 1]}, 1)$.
	\ENDIF
	\FOR{MCMC iteration $t=1$ to $T$}
	\STATE Sample $\hat{\boldsymbol{\gamma}}_{\text{pot}}^{(t)}$ from the multivariate normal distribution $N(\boldsymbol{\mu}_{\boldsymbol{\gamma}}, \boldsymbol{\Sigma}_{\boldsymbol{\gamma}})$ with
		\begin{align*}
		\boldsymbol{\Sigma}_{\boldsymbol{\gamma}} &= \left(\boldsymbol{Z}^T \boldsymbol{R}^{-1} \boldsymbol{Z} + \boldsymbol{G}^{-1}\right)^{-1} \\
		\boldsymbol{\mu}_{\boldsymbol{\gamma}} &= \boldsymbol{\Sigma}_{\boldsymbol{\gamma}} \left(\boldsymbol{Z}^T \boldsymbol{R}^{-1} \left(\boldsymbol{y} - \boldsymbol{X} \hat{\boldsymbol{\beta}}^{[s]}\right)\right),
		\end{align*}
	where $\boldsymbol{Z} = \boldsymbol{Z}_{\text{pot}}^{[s]}$, $\boldsymbol{R} = \boldsymbol{R}^{(t-1)}$ and $\boldsymbol{G} = \boldsymbol{G}^{(t-1)}$.
	\STATE Sample $\hat{\sigma}^{2(t)}$ from the inverse Gamma distribution $IG(\tilde{a}, \tilde{b})$ with
		\begin{align*}
		\tilde{a} &= a + \frac{n}{2} \\
		\tilde{b} &= b + \frac{1}{2}\left(\boldsymbol{y} - \boldsymbol{X}\hat{\boldsymbol{\beta}}^{[s]} - \boldsymbol{Z}\boldsymbol{\gamma}\right)^T \left(\boldsymbol{y} - \boldsymbol{X}\hat{\boldsymbol{\beta}}^{[s]} - \boldsymbol{Z}\boldsymbol{\gamma}\right),
		\end{align*}
	where $\boldsymbol{\gamma} = \hat{\boldsymbol{\gamma}}_{\text{pot}}^{(t)}$. Then set $\boldsymbol{R}^{(t)} = \hat{\sigma}^{2(t)}\boldsymbol{I}$.
	\STATE Sample $\hat{\boldsymbol{Q}}_{\text{pot}}^{(t)}$ from the inverse Wishart distribution $IW(v, \boldsymbol{\Lambda})$ with
		\begin{align*}
		v &= v_0 + m \\
		\boldsymbol{\Lambda} &= \boldsymbol{\Lambda}_{0, \text{pot}}^{[s]} + \boldsymbol{\gamma}^T\boldsymbol{\gamma},
		\end{align*}
	where $\boldsymbol{\gamma} = \boldsymbol{\gamma}_{\text{pot}}^{(t)}$. Then set $\boldsymbol{G}^{(t)} = \text{blockdiag}(\hat{\boldsymbol{Q}}_{\text{pot}}^{(t)}, \cdots, \hat{\boldsymbol{Q}}_{\text{pot}}^{(t)}, \cdots, \hat{\boldsymbol{Q}}_{\text{pot}}^{(t)})$.
	\ENDFOR
	
	\STATE Compute the modes $\hat{\boldsymbol{\gamma}}_{\text{mode, pot}}^{[s]}$ from elementwise $\hat{\boldsymbol{\gamma}}_{\text{pot}}^{(t)}$ sample distributions.
	\STATE Compute the mode $\hat{\sigma}_{\text{mode}}^{2[s]}$ from $\hat{\sigma}^{2(t)}$ sample distribution,
	and set $\sigma^{2(0)} = \sigma_{\text{mode}}^{2[s]}$, therefore, $\boldsymbol{R}^{(0)} = \sigma_{\text{mode}}^{2[s]} \boldsymbol{I}$.
	\STATE Compute the modes $\hat{\boldsymbol{Q}}_{\text{mode, pot}}^{[s]}$ from elementwise $\hat{\boldsymbol{Q}}^{(t)}$ sample distributions.
	\STATE Calculate the mean squared error of the random effects part,
		\begin{align*}
		\text{MSE}_{k^*} = \frac{1}{n} \sum_{i = 1}^n (u_i - \boldsymbol{X}_{ik^*} \hat{\boldsymbol{\beta}}_{k^*}^{[s]} - \boldsymbol{Z}_{ik^*, \text{pot}}^{[s]} \boldsymbol{\gamma}_{k^*, \text{mode, pot}}^{[s]})^2.
		\end{align*}
	\IF{$\min(\text{MSE}_{k^*, \text{fixed}}) \leq \min(\text{MSE}_{k^*})$}
	\STATE Reject the potential structure and reset them to the state in the last boosting iteration
		\begin{align*}
		\boldsymbol{Z}_{\text{pot}}^{[s]} &= \boldsymbol{Z}^{[s - 1]} \\
		\boldsymbol{Z}^{[s]} &= \boldsymbol{Z}_{\text{pot}}^{[s - 1]} \\
		\boldsymbol{\Lambda}_{0, \text{pot}}^{[s]} &= \boldsymbol{\Lambda}_{0}^{[s - 1]}.
		\end{align*}
	\ELSE
	\STATE Select $X_{k^*}$ as an informative random effect and accept the potential structure,
		\begin{align*}
		E &= E \cup X_{k^*} \\
		\boldsymbol{Z}^{[s]} &= \boldsymbol{Z}_{\text{pot}}^{[s]} \\
		\boldsymbol{\Lambda}_0^{[s]} &= \boldsymbol{\Lambda}_{0,  \text{pot}}^{[s]}.
		\end{align*}
	\ENDIF
	\STATE Subset estimates to contain only set $E$ relevant covariates,
		\begin{align*}
		\hat{\boldsymbol{\gamma}}_{\text{mode}}^{[s]} &= \boldsymbol{\gamma}_{\text{mode, pot}}^{[s]} \cap E \\
		\boldsymbol{Q}_{\text{mode}}^{[s]} &= \boldsymbol{Q}_{\text{mode, pot}}^{[s]} \cap E,
		\end{align*}
	\STATE Update
		\begin{align*}
		\hat{\boldsymbol{y}}^{[s]} = \boldsymbol{X}\hat{\boldsymbol{\beta}}^{[s]} + \boldsymbol{Z}^{[s]}\hat{\boldsymbol{\gamma}}_{\text{mode}}^{[s]}.
		\end{align*}
\ENDFOR
\end{algorithmic}
\end{breakablealgorithm}

\end{document}